\documentclass{article}
\usepackage [utf8] {inputenc}
\usepackage{mathtools}
\usepackage{amsthm}
\usepackage{amssymb}
\usepackage{enumerate}
\usepackage{hyperref}

\newtheorem{theorem}{Theorem}

\DeclareMathOperator{\Aut}{Aut}
\DeclareMathOperator{\Hom}{Hom}

\title{Reduction of the group isomorphism problem to the group automorphism problem}
\author{Saveliy V. Skresanov}
\date{}

\begin{document}
\maketitle

\begin{abstract}
	It is well known that the graph isomorphism problem is polynomial-time
	reducible to the graph automorphism problem (in fact these two problems
	are polynomial-time equivalent). We show that, analogously, the group isomorphism problem
	is polynomial-time reducible to the group automorphism problem. Reductions to other relevant
	problems like automorphism counting are also given.
\end{abstract}

\section{Introduction}

The \emph{graph isomorphism problem}~--- the problem of deciding whether two input graphs given by their adjacency matrices
are isomorphic or not~--- is one of the main candidates for a problem of intermediate computational complexity.
Indeed, graph isomorphism can be easily seen to lie in class NP, and the common conjecture is that it is not NP-complete,
see some evidence in~\cite{schoning}. Nevertheless, no polynomial-time algorithm is known for graph isomorphism.

The long history of improving time complexity of graph isomorphism algorithms culminated in
a 2015 breakthrough by Babai~\cite{babai}, who gave a quasipolynomial algorithm for that task.
More precisely, his algorithm decides whether two graphs on \( n \) vertices are isomorphic or not in time \( n^{O((\log n)^c)} \)
for some constant \( c \geq 1 \). This brings time complexity of graph isomorphism close to the one for the \emph{group isomorphism problem},
that is, the problem of deciding whether two groups given by their Cayley tables are isomorphic or not.
Indeed, the latter has a simple algorithm attributed to Tarjan in~\cite{miller} of complexity \( n^{O(\log n)} \), where \( n \) is the order of the input groups.

One important property of the graph isomorphism problem often used in studying its complexity, is the fact that it is
polynomial-time equivalent to many other related problems, see~\cite{mathon}. For example, it is polynomial-time equivalent
to the problem of finding generators of the full automorphism group of some input graph.
Since Luks's algorithm for isomorphism of graphs of bounded degree~\cite{luks}, this latter problem played
the main role in enabling the use of group-theoretical techniques in isomorphism testing.

The reduction of the graph isomorphism problem to the problem of finding the full automorphism group of some graph is easy to state.
By considering connected components, we may assume that our input graphs \( \Gamma_1 \) and \( \Gamma_2 \) are connected.
Compute generators of the full automorphism group of their disjoint union \( \Gamma_1 \sqcup \Gamma_2 \). Then
\( \Gamma_1 \) and \( \Gamma_2 \) are isomorphic if and only if there exists a generator which swaps their vertex sets.
Given access to the graph automorphism oracle, it can be easily seen that all the required steps can be computed in polynomial time.

The goal of this paper is to provide a similar reduction of the group isomorphism problem to the group automorphism problem.
The idea of the proof follows the procedure described above, but instead of connected components we consider directly indecomposable groups,
and direct products instead of disjoint unions. To make the argument work, we require some nontrivial results and algorithms for finite groups,
for example, the description of the full automorphism group of a direct product by Bidwell, Curran and McCaughan~\cite{bidwell, bidwell2},
and a polynomial-time algorithm for factoring groups into direct factors by Kayal and Nezhmetdinov~\cite{kayal}.

To state the result, we introduce a set of computational problems for groups, which follow similar problems for graphs from~\cite{mathon}.
Recall that all input groups are given by their Cayley tables, and generators of the full automorphism group \( \Aut(G) \) of a group \( G \)
are given as permutations on the set \( G \). The automorphic partition of \( G \) is the set of orbits of \( \Aut(G) \)
in the natural action on~\( G \).
\begin{align*}
	&\mathrm{GrpISO}(G, H): \text{ determine if two groups } G \text{ and } H \text{ are isomorphic or not.}\\
	&\mathrm{GrpIMAP}(G, H): \text{ find an isomorphism between } G \text{ and } H \text{ if it exists.}\\
	&\mathrm{GrpICOUNT}(G, H): \text{ count the number of isomorphisms between } G \text{ and } H.\\
	&\mathrm{GrpACOUNT}(G): \text{ count the number of automorphisms of } G.\\
	&\mathrm{GrpAGEN}(G): \text{ find generators of the full automorphism group of } G.\\
	&\mathrm{GrpAPART}(G): \text{ find the automorphic partition of } G.
\end{align*}

For two computational problems \( P_1 \) and \( P_2 \) we write \( P_1 \propto P_2 \) if \( P_1 \) is polynomial-time reducible to \( P_2 \).
Problems \( P_1 \) and \( P_2 \) are polynomial-time equivalent if \( P_1 \propto P_2 \) and \( P_2 \propto P_1 \).

\begin{theorem}
	The following hold:
	\begin{enumerate}[\rm(i)]
		\item \( \mathrm{GrpISO} \propto \mathrm{GrpACOUNT} \propto \mathrm{GrpAGEN} \),
		\item \( \mathrm{GrpISO} \propto \mathrm{GrpAPART} \propto \mathrm{GrpAGEN} \),
		\item \( \mathrm{GrpIMAP} \propto \mathrm{GrpAGEN} \),
		\item \( \mathrm{GrpICOUNT} \) and \( \mathrm{GrpACOUNT} \) are polynomial-time equivalent.
	\end{enumerate}
\end{theorem}

We noted that for graphs, isomorphism and automorphism problems are polynomial-time equivalent.
The reduction of automorphism group computation to isomorphism testing uses graph gadgets, which essentially allow coloring
graph vertices. By individualising graph vertices in a special way, it is possible to deduce automorphisms of the graph
using several queries to the graph isomorphism oracle. Unfortunately this technique seems to be unavailable
for groups, as it is not clear how to ``color'' group elements while preserving the group structure.

One way to remedy this is to consider isomorphism problems for \emph{colored groups}, where elements of the groups can be colored
and isomorphisms must preserve the colors. It is shown in~\cite[Lemma~8.1]{chenpon} that problems \( \mathrm{GrpISO} \), \( \mathrm{GrpIMAP} \)
and \( \mathrm{GrpAGEN} \) are polynomial-time equivalent for colored groups, and in fact the proof of~\cite[Lemma~8.1]{chenpon} can be easily adapted
to show that problems \( \mathrm{GrpICOUNT} \), \( \mathrm{GrpACOUNT} \) and \( \mathrm{GrpAPART} \) are also polynomial-time equivalent to \( \mathrm{GrpISO} \)
for colored groups. We remark that even for abelian colored groups the isomorphism problem is highly nontrivial, since a polynomial-time algorithm
for such a problem would greatly simplify current algorithms for computing 2-closures of primitive affine permutation groups.
Indeed, in many important cases the latter problem reduces to the following task: given a finite vector space \( V \) and a linear group \( G \leq \mathrm{GL}(V) \)
acting on \( V \), find the largest subgroup of \( \mathrm{GL}(V) \) having the same orbits on \( V \) as~\( G \).
This can be easily seen to be reducible to the problem of computing the full automorphism group of a colored abelian group \( V \),
where the color classes are the orbits of \( G \) on~\( V \).
Current algorithms for this problem heavily utilize the group-theoretical structure of \( G \) and do not work for all input groups,
for example, see~\cite{rank3} and references therein.

\section{Preliminaries}

We refer the reader to~\cite{gorenstein} for the basic facts about finite groups.
We write \( G \times H \) for the direct product of groups \( G \) and \( H \), and \( G \rtimes H \)
for the semidirect product, where \( H \) is assumed to act on \( G \) by automorphisms.
Let \( Z(G) \) denote the center of \( G \), and let \( [G, G] \) stand for the derived subgroup.
We write \( \Aut(G) \) for the full automorphism group of \( G \), and \( \Hom(G, H) \)
for the set of all homomorphisms from the group \( G \) into the group \( H \).
Given an element \( g \in G \), we write \( |g| \) for its order.

Input groups to the algorithms are given by their Cayley tables. A subgroup of such a group
is given as a list of its elements, and a homomorphism \( {\phi : G \to H} \) is given as a list of pairs \( (g, \phi(g)) \), \( g \in G \);
observe that the space required is polynomial in the sizes of the input groups.
Note that standard group-theoretical tasks such as computing direct products, quotient groups, finding the center and derived subgroup
can be performed in polynomial time for groups given by Cayley tables.

\section{Proof of the main result}
The order of a permutation group given by generators and its orbits on the permutation domain can be found in polynomial time, see~\cite{seress},
so \( \mathrm{GrpACOUNT} \) and \( \mathrm{GrpAPART} \) are polynomial-time reducible to \( \mathrm{GrpAGEN} \).

Let \( G \) and \( H \) be two groups which we may assume are both of order~\( n \).
Using~\cite{kayal}, in polynomial time we can decompose our input groups as \( G \simeq G_1 \times \cdots \times G_k \)
and \( H \simeq H_1 \times \cdots \times H_m \), where \( G_i \) and \( H_j \) are directly indecomposable.
By the Krull--Shmidt theorem, these decompositions are unique up to reordering of the factors, hence to check isomorphism between \( G \) and \( H \)
it suffices to check isomorphism between \(k \cdot m \) pairs of groups \( G_i, H_j \). Since \( k, m \leq \log n \),
\( \mathrm{GrpISO} \) is polynomial-time reducible to the isomorphism problem of directly indecomposable groups.
Similarly, we may assume that groups are directly indecomposable in \( \mathrm{GrpIMAP} \).

Now, let \( G \) and \( H \) be directly indecomposable.
Given two maps \( \theta : B \to C \) and \( \phi : A \to B \) for groups \( A, B, C \), define their product \( \theta \cdot \phi : A \to C \)
as the composition \( (\theta \cdot \phi)(a) = \theta(\phi(a)) \), \( a \in A \). Given maps \( \theta, \phi : A \to B \), define their sum
as the pointwise product \( (\theta + \phi)(a) = \theta(a)\phi(a) \). We define the following group of formal matrices:
\[ \mathcal{A} = \left\{ \begin{pmatrix} \alpha & \beta \\ \gamma & \delta \end{pmatrix} \mid \begin{aligned} &\alpha \in \Aut(G), \beta \in \Hom(H, Z(G)),\\
&\gamma \in \Hom(G, Z(H)), \delta \in \Aut(H) \end{aligned} \right\}, \]
where the group operation is matrix multiplication, and products and sums of maps are computed as described above.
If we represent an element of \( G \times H \) as a tuple \( (g, h) \), \( g \in G \), \( h \in H \), then an element of \( \mathcal{A} \)
acts on it by multiplying the column vector \( (g, h)^T \) from the left by the corresponding matrix:
\[ \tag{\(\star\)} \begin{pmatrix} \alpha & \beta \\ \gamma & \delta \end{pmatrix} (g, h)^T = (\alpha(g)\beta(h), \gamma(g)\delta(h))^T. \]
It is proved in~\cite{bidwell, bidwell2} that if \( G \) and \( H \) are not isomorphic, then \( \Aut(G \times H) \simeq \mathcal{A} \),
and if \( G \) and \( H \) are isomorphic, then \( \Aut(G \times H) \simeq \mathcal{A} \rtimes C_2 \),
where \( C_2 \) is the group of order~2 acting on \( G \times H \) by swapping the direct factors.

First we prove part~(i), namely, \( \mathrm{GrpISO} \propto \mathrm{GrpACOUNT} \). Note that
\[ |\Aut(G \times H)| = |\Aut(G)| \cdot |\Aut(H)| \cdot |\Hom(G, Z(H))| \cdot |\Hom(H, Z(G))| \cdot \epsilon_{G,H}, \]
where \( \epsilon_{G,H} = 1 \) if \( G \not\simeq H \) and \( \epsilon_{G,H} = 2 \) otherwise.
By using three calls to the \( \mathrm{GrpACOUNT} \) oracle, we can compute \( |\Aut(G \times H)| \), \( |\Aut(G)| \) and \( |\Aut(H)| \).
Hence to determine whether \( G \simeq H \) or not, it suffices to compute \( |\Hom(G, Z(H))| \) and \( |\Hom(H, Z(G))| \) in polynomial time.

Recall that every homomorphism from a group into an abelian subgroup contains the derived subgroup in its kernel.
Hence
\begin{align*}
	|\Hom(G, Z(H))| &= |\Hom(G/[G,G], Z(H))|,\\
	|\Hom(H, Z(G))| &= |\Hom(H/[H,H], Z(G))|, 
\end{align*}
so the problem reduces to counting the number of homomorphisms between two abelian groups \( A \) and \( B \). If \( A \simeq A_1 \times A_2 \) where
\( |A_1| \) and \( |A_2| \) are coprime, then \( |\Hom(A, B)| = |\Hom(A_1, B)| \cdot |\Hom(A_2, B)| \), so we may assume that \( A \)
is a group of prime power order. Decompose \( A \) into a direct product of cyclic groups with generators \( a_1, \dots, a_r \),
where \( r \) is the rank of \( A \). For an integer \( t \) let
\[ N(t) = |\{ b \in B \mid b^t = 1 \}| \]
be the number of elements of order dividing \( t \) in \( B \). If \( \phi \in \Hom(A, B) \),
then \( \phi \) is completely determined by elements \( \phi(a_1), \dots, \phi(a_r) \in B \).
Clearly \( \phi(a_i)^{|a_i|} = 1 \) for all \( i \), and for any \( b_1, \dots, b_r \in B \)
with \( b_i^{|a_i|} = 1 \), \( i = 1, \dots, r \), there exists a unique \( \phi \in \Hom(A, B) \)
with \( \phi(a_i) = b_i \), \( i = 1, \dots, r \), defined by the formula
\[ \phi(a_1^{k_1} \cdots a_r^{k_r}) = b_1^{k_1} \cdots b_r^{k_r}, \]
for all integers \( k_1, \dots, k_r \).
Hence \( |\Hom(A, B)| = N(|a_1|) \cdot \dots \cdot N(|a_r|) \).
Since these computations are easily performed in polynomial time, we can compute \( |\Hom(G/[G,G], Z(H))| \) and \( |\Hom(H/[H,H], Z(G))| \)
thus finding \( \epsilon_{G,H} \) in polynomial time. Part~(i) is proved.

Now we prove part~(ii), the reduction \( \mathrm{GrpISO} \propto \mathrm{GrpAPART} \).
We may assume that \( G \) and \( H \) are not abelian, since the isomorphism problem for abelian groups can be solved in polynomial time~\cite{lipton}.
Consider a subgroup \( S = G \times Z(H) \) of \( G \times H \). If \( G \) and \( H \) are not isomorphic, then we claim that \( S \) is
invariant under all automorphisms of \( G \times H \). Indeed, let \( (g, h) \in S \) be an arbitrary element. Since \( \Aut(G \times H) \simeq \mathcal{A} \),
it suffices to check the action of an arbitrary element of \( \mathcal{A} \) on \( (g, h) \). In the notation of formula~(\(\star\)),
observe that \( \alpha(g)\beta(h) \in G \), while \( \gamma(g) \in Z(H) \) since \( \gamma \in \Hom(G, Z(H)) \), and \( \delta(h) \in Z(H) \)
since \( h \in Z(H) \) and \( \delta \in \Aut(H) \). Therefore \( (\alpha(g)\beta(h), \gamma(g)\delta(h)) \in S \) as claimed.

If \( G \simeq H \), then recall that \( \Aut(G \times H) \) contains an additional automorphism flipping direct factors. Let \( \pi : G \to H \) be an isomorphism,
and define \( \phi \in \Aut(G \times H) \) by \( \phi((g, h)) = (\pi^{-1}(h), \pi(g)) \), where \( (g, h) \in G \times H \).
Now, take \( (g, h) \in S \) such that \( g \not\in Z(G) \); this is possible since \( G \) is nonabelian.
Then \( \pi(g) \not\in \pi(Z(G)) = Z(H) \), hence \( \phi((g, h)) \not\in S \). To sum up, the subgroup \( S \) is invariant under all automorphisms
of \( \Aut(G \times H) \) if and only if \( G \) is not isomorphic to \( H \).

Using one call to the \( \mathrm{GrpAPART} \) oracle we can compute the automorphic partition of \( G \times H \).
The subgroup \( S \) will be invariant under all automorphisms if and only if it is a union of some parts of this partition,
and this can be tested in polynomial time. Part~(ii) is proved.

We prove part~(iii). Using one call to the \( \mathrm{GrpAGEN} \) oracle we compute generators of \( \Aut(G \times H) \).
We can check in polynomial time whether all generators stabilize the set \( S \). If they do, then \( G \) is not isomorphic to \( H \),
and we output nothing. If there exists a generator \( \psi \in \Aut(G \times H) \) such that \( \psi(S) \neq S \),
then \( G \simeq H \). Let \( \pi : G \to H \) be some isomorphism, and define \( \phi \in \Aut(G \times H) \)
by \( \phi((g, h)) = (\pi^{-1}(h), \pi(g)) \), \( g \in G \), \( h \in H \). The group \( \Aut(G \times H) \) decomposes
into a semidirect product of \( \mathcal{A} \) and the group of order two generated by \( \phi \),
hence \( \psi(x) = \phi(\theta(x)) \) for some \( \theta \in \mathcal{A} \).
Consider the action of \( \psi \) on \( (g, 1) \in G \times H \). In the notation of formula~\((\star)\) the automorphism \( \theta \) acts by
\( \theta((g, 1)) = (\alpha(g)\beta(1), \gamma(g)\delta(1)) = (\alpha(g), \gamma(g)) \). Hence
\[ \psi((g, 1)) = \phi(\theta((g,1))) = \phi((\alpha(g), \gamma(g))) = (\pi^{-1}(\gamma(g)), \pi(\alpha(g))). \]
Since \( \alpha \in \Aut(G) \), the map \( g \mapsto \pi(\alpha(g)) \) is an isomorphism between \( G \) and \( H \),
and we can find it in polynomial time by computing \( \psi((g, 1)) \in G \times H \) and taking the projection of the result to \( H \).
Part~(iii) is proved.

Finally, to prove part~(iv), note that \( \mathrm{GrpACOUNT} \propto \mathrm{GrpICOUNT} \) holds trivially,
since \( |\Aut(G)| \) is equal to the number of isomorphisms between \( G \) and itself. To show \( \mathrm{GrpICOUNT} \propto \mathrm{GrpACOUNT} \),
let \( G \) and \( H \) be two input groups. Since \( \mathrm{GrpISO} \propto \mathrm{GrpACOUNT} \) by part~(i),
we can test whether \( G \simeq H \) in polynomial time.
If \( G \not\simeq H \), then the number of isomorphisms between \( G \) and \( H \) is zero. If \( G \simeq H \),
then the number of isomorphisms between \( G \) and \( H \) is \( |\Aut(G)| \), which can be computed with one call to the \( \mathrm{GrpACOUNT} \) oracle.
Part~(iv) is proved. \qed

\section{Acknowledgements}

The author expresses his gratitude to I.N.~Ponomarenko and A.V.~Vasil'ev for helpful comments improving the text.

The work is supported by the Russian Science Foundation, project no.~24-11-00127, \url{https://rscf.ru/en/project/24-11-00127/}.

\bigskip

\noindent
\emph{Saveliy V. Skresanov}

\noindent
\emph{Novosibirsk State University, 1 Pirogova St.,}

\noindent
\emph{Sobolev Institute of Mathematics, 4 Acad. Koptyug avenue,\\ 630090 Novosibirsk, Russia}

\noindent
\emph{Email address: skresan@math.nsc.ru}

\end{document}